\documentstyle[preprint,aps]{revtex}

\begin{document}
\draft
\title{Off-shell effects in heavy particle production}
\author{G.F.Bertsch$^{a}$ and P. Danielewicz$^{b}$}
\address{$^{a}$Physics Department and Institute for Nuclear
Theory\\
University of Washington, Seattle, WA 98195\\ and\\
$^b$Department of Physics and Astronomy and National
Superconducting Cyclotron Laboratory\\ Michigan State
University, East Lansing, MI 48824}
\maketitle

\begin{abstract}
Off-shell propagation of nucleons is neglected in
one-body transport models of heavy-ion collisions, but it
could be significant in processes that are limited by phase
space, such as the threshold production of heavy particles.
We estimate the relative magnitude of off-shell production to
on-shell production of the N$^*$(1535) resonance in heavy ion
collisions.   In the region where the on-shell production is
dominated by a~two-step mechanism with an intermediate $\Delta$,
we find that the contribution of off-shell scattering between
projectile and target
nucleons is indeed small.  Beyond the latter contribution,
however, correlations in
the initial wave function produce off-shell contributions which
can exceed those of the on-shell $\Delta$ mechanism.

\end{abstract}


One-body transport theory is widely applied to interpret heavy-ion
reactions.  In the usual application of the theory, such as
with the Boltzmann-Uehling-Uhlenbeck (BUU)
equation\cite{BD88}, one assumes
that particles propagate classically between collisions.
A~more complete theory would allow off-shell propagation of particles;
the validity of BUU and similar theories depends on these
off-shell effects being small.

Recently the claim has been advanced that near-threshold
heavy-particle production in nucleus-nucleus collisions is
mediated by intermediate production of delta
resonances\cite{bat91,wol93,gqli92,har94,met93,bali94}.
The conclusion was based on studies within the BUU theory,
and thus depends on off-shell propagation being insignificant.

Our objective in this letter is to estimate off-shell effects
for such processes.  We will make several drastic simplifying
assumptions, in order to derive simple formulas. First,
we~shall consider the stationary rate of production from the
initial
distribution of nucleons in the colliding nuclei.  We also
neglect the Fermi momentum of the nucleons, assuming that in
zeroth order they are in momentum eigenstates at the beam and
target velocities.  The~sensitivity to off-shell effects should
be highest below the energy threshold for the heavy particle
in a two-body collision.  At such energies the lowest
graph that contributes involves three particles, as shown
in~Fig.~1.  In the graph, the particle labeled by $k_3$ is
a~baryon resonance such as the $N^*(1535)$.
The intermediate particle, labeled by~R, could
be a~delta resonance or a~nucleon.  The kinematics is
such that
a nucleon always propagates off its energy shell; the
delta may
be on shell or not depending on the momenta in the final state.

The two interactions will be much simplified; at the moment we
denote the two matrix elements as  $M_1$ and $M_2$.
Then the amplitude to
go from an initial state with momenta $(p_1,p_2,p_3)$ to a
final state with momenta $(k_1,k_2,k_3)$ is
\begin{equation}
{M_1M_2 \over  \Delta E +i\Gamma/2}
\label{amplitude}
\end{equation}
where $\Delta E = E(p_1) + E(p_2)
-E(k_1)-E_{R}(p_1+p_2-k_1)$ and $E(k)=(m^2+k^2)^{1/2}$.
The $\Gamma$ in the propagator is the width of the intermediate
resonance in the
medium.  The rate for the reaction is given by the
modulus squared of the amplitude~(\ref{amplitude}) summed  over
final states and multiplied by the initial nucleon densities,
\begin{eqnarray}
W & = & n_1 \, n_2 \, n_3 \int {d^3 k_1 \over (2\pi)^3} \, {
d^3 k_2\over (2\pi)^3} \, {  d^3 k_3\over (2\pi)^3} \,
|M_1|^2 \, {1\over (\Delta E )^2+\Gamma^2/4} \, |M_2|^2
\nonumber\\[.1cm]
&& \times
(2\pi)^4\delta^{(4)}(k_1+k_2+k_3- p_1 -p_2-p_3).
\label{W=}
\end{eqnarray}
It is convenient to write this in the factorized form,
\begin{eqnarray}
W & = & n_1 \, n_2 \int {d^3 k_1 \over (2\pi)^3} \, {
d^4 k_R\over (2\pi)^4} \,
|M_1|^2 \, {\Gamma \over (\Delta E )^2+\Gamma^2/4} \,
(2\pi)^4\delta^{(4)}(k_1+k_R- p_1 -p_2)
\nonumber\\[.1cm]
&& \times {1 \over \Gamma} \, n_3 \int {  d^3 k_2\over
(2\pi)^3} \,
 {  d^3 k_3\over
(2\pi)^3} \,
|M_1|^2 \,
(2\pi)^4\delta^{(4)}(k_2+k_3- k_R-p_3)
\nonumber\\[.1cm]
&  = & \int d^4 k_R \, {d^4 W_{NN \rightarrow NR} \over
d k_R^4 } \, {\Gamma_{NR \rightarrow NN^*} \over \Gamma}
\, ,
\label{Wf}
\end{eqnarray}
where we have replaced the $\delta^{(4)}$-function
in~(\ref{W=}) by a~convolution of two functions.  In the
last line we have expressed the rate as the product
of a rate for producing a~generally off-shell particle,
$W_{NN \rightarrow NR}$, times the branching ratio for that
particle to produce~$N^*$ before disappearing from the system,
$\Gamma_{NR \rightarrow NN^*}/ \Gamma$.  Beyond
this, $\Gamma^{-1} \, d^4 W_{NN \rightarrow NR} / d k_R^4$
in~(\ref{Wf}) represents a~contribution of the~$NN$
interactions to
the number of off-shell particles~$R$ per unit phase-space and
energy at any instant.
We will return to this interpretation
in the conclusion.

When a delta is produced in the intermediate state of the
reaction and propagates on shell,  the integral for the rate is
dominated by the region of the integrand with $\Delta E \sim
\Gamma$ due to the factor $((\Delta E
)^2+\Gamma^2/4)^{-1}$. Assuming
that all other factors in the integrand are constant over this
range,
the integral over
$k_R^0$ can be evaluated using
\begin{equation}
\int {d k_R^0 \over 2\pi} {\Gamma \over ( k_R^0
-E_R(k_r))^2+\Gamma^2/4} = 1
\end{equation}
The result is
\begin{eqnarray}
W_\Delta & = & n_1 \, n_2 \int {d^3 k_1\over (2\pi)^3} \, {d^3
k_\Delta \over (2\pi)^3} \,
|M_1|^2 \, (2\pi)^4 \delta^{(4)}(k_1+k_\Delta-p_1-p_2)
\nonumber\\[.1cm]
& &  \times {1 \over \Gamma_\Delta} \, n_3  \int {d^3 k_2\over
(2\pi)^3} \, {d^3 k_3\over (2\pi)^3} \, |M_2|^2 \,
(2\pi)^4 \delta^{(4)}(k_\Delta+p_3-k_2-k_3)
\nonumber\\[.1cm]
& = & \int d \Omega_{N\Delta} \, {d W_{NN\rightarrow N\Delta}
\over d \Omega_{N\Delta}}
{ \Gamma_{N\Delta\rightarrow NN^*}  \over \Gamma_\Delta}  \, .
\label{WD}
\end{eqnarray}
Here  $W_{NN\rightarrow N\Delta}$ is the physical rate for
producing on-shell deltas in the medium.

When the off-shell propagation dominates, we can neglect
$\Gamma$ in the denominator in~Eq.~(\ref{W=}).  The~integration
over momenta in Eq.~(\ref{W=}) is conveniently
done by going first to the cm frame of particles 2' and 3' in
the final state,
which allows an angular integration to be done
trivially.
The~integral that remains  to be evaluated for the off-shell
rate is then
\begin{equation}
W_{off} =  n_1 \, n_2 \, n_3 \, {4\pi\over(2\pi)^5}
\int_0^{k_{max}} dk_1 \int d\Omega_{k_1}  \,  k_1^2  \, k_2
\, \mu_{k_2} \, {|M_1M_2|^2\over(\Delta E)^2}.
\label{Woff}
\end{equation}
Here $\mu_{k_2}=(1/\sqrt{m_{N}^2
+ k_2^2}+1/\sqrt{m_{N^*}^2+k_2^2})^{-1}$ is the reduced
mass associated with $k_2$, $k_1$~is the momentum in
3-body cm frame, and~the
limit $k_{max}$ is determined by the threshold where $k_2=0$.

The interesting quantity for our purposes is the ratio of the
rates
by the two mechanisms.  We reduce Eq.~(\ref{WD}) in the similar
way we derived Eq.~(\ref{Woff}), obtaining
\begin{equation}
W_\Delta = {n_1 \, n_2 \, n_3 \over \Gamma_\Delta} {4\pi\over
(2\pi)^4}\int
d \Omega_{k_1} \, k_1 \, \mu_{k_1} \, k_2 \,
\mu_{k_2} \, |M_1M_2|^2 .
\label{WDi}
\end{equation}
Here $\vec{k_1}$ is in the $N\Delta$ cm frame.
Assuming now that
the variation of elements
$M_{1,2}$ is small on the scale of the integrations,
the ratio of the two rates becomes
\begin{equation}
{W_{off}\over W_\Delta}={\Gamma_\Delta \over 2\pi}
{\int_0^{k_{max}} dk_1 \,  d \Omega_{k_2}\, k_1^2 \, k_2
\mu_{k_2} / (\Delta E)^2
\over \int d\Omega_{k_1} \, k_1 \, \mu_{k_1} \, k_2  \,
\mu_{k_2} } \, .
\label{Wr}
\end{equation}
The r.h.s.~of
Eq.~(\ref{Wr}) depends only on $\Gamma$ and kinematic
quantities; the dependence
on matrix elements is canceled out by our assumption that they
are the same for nucleons as for deltas.

We are now ready to evaluate numerically the relative
contribution of the two mechanisms.  We choose an initial
beam momentum of $p_b=1.9$ GeV/c, which is below
the two-body threshold at
2.1 GeV/c, but above both three-body thresholds.
We~take the mass and width of the delta to be
$m_\Delta=1.23$~GeV and $\Gamma_\Delta = 0.12$~GeV.
First, let us consider first the situation where particle~1 is
from the beam and the other two are in the target,
i.e.~$(p_1,p_2,p_3)=(p_b,0,0)$.  For the delta mechanism,
the first scattering (between~1 and~2) produces a delta
whose momentum in
the $N\Delta$
cm system is $k_\Delta= 0.5$ GeV/c.  There is enough energy to
produce an N$^*$(1535) only if the delta goes forward;
the solid angle integration in
Eq.~(\ref{WDi}) extends only over $\Omega_\Delta \,
\raisebox{-.5ex}{$\stackrel{<}{\scriptstyle\sim}$} \, 0.2 \times
4 \pi$.  The~approximate magnitudes of the other quantities in
this angular range are $k_2\approx 0.2$~GeV/c and $\mu \sim
m_N/2$.  For the off-shell mechanism, there is no kinematic
restriction on the intermediate state and the full range of
solid angle contributes.  The range of the $k_1$ integration
is determined by the available energy to produce the N$^*$(1535);
for our conditions $k_{max}\approx0.4$~GeV/c in the 3-body
cm frame.  The remaining quantity to include is $\Delta E\approx
- 0.5$~GeV/c.  One can estimate the integrals with these
numbers to get a ratio   $W_{off}/W_\Delta \approx 15\%$.

{}From this result the on-shell theory looks reasonably
accurate, but
we still have to examine other combinations of projectile
and target particle in the basic graph in Fig.~1.  It~is
sufficiently
general to consider one particle from the projectile and two
from the target, but we need also to consider the case where~1
and~2 are in the same nucleus, i.e.~$(p_1,p_2,p_3)=(0,0,p_b)$.
Now the first interaction represents an~initial-state
correlation.
It obviously cannot make an on-shell delta, but it can give
a~nucleon the momentum needed to produce a~heavy particle
in the second interaction.
The energy denominator is small in
the kinematic region where~$\vec{k_1}$ is maximal and in the
forward direction, reaching $\Delta E=- 0.01$~GeV at the limit.
However, phase space favors larger values of off-shell energy
with phase-space average being
$\langle \Delta E \rangle \approx - 0.1$~GeV.  The~ratio of
integrals with these numbers is~$W_{corr}/W_\Delta \approx 3$.

Thus we find that the mechanism involving initial state
correlations will dominate
the threshold production of heavy particles, at least
early in a~reaction.  Under this mechanism, the target
correlations cause a target nucleon
to acquire a large backward momentum before it interacts with
a~projectile nucleon.
In~the example above, the momenta of the
nucleon moving backwards in the target frame range from
0.1~GeV/c to~1~GeV/c.  While this nucleon is highly
off-shell,
overall enough energy is gained in the two-nucleon frame of
two nucleons to allow the final heavy-particle production.
Generally, because of the presence of energetic nucleons from
the other nucleus, the~high momenta from correlations can
play a~role and materialize in interactions.
When $p_1$ and $p_2$ are from the same nucleus, then
$\Gamma^{-1} \, d^4 W_{NN \rightarrow NN} / d k^4$
in~(\ref{Wf}) represents simply a~contribution of $NN$
interactions to the ground-state nucleon spectral-function.
When the heavy particle production
is to take place deeply below  the $NN$ threshold,
the~correlations involving just two target nucleons may not suffice
to generate the necessary energy in the three-body cm frame.
It~may be then advantageous to involve more nucleons in the
initial-state correlation.
The~view
that correlations dominate threshold production has been
implicit in calculations of Ref.\cite{dan90} and the effects of
two- and more-particle correlations have been included there.
As~a~heavy-ion reaction progresses, the~balance in the rates
may shift from the mechanism involving correlations to the
mechanism
with on-shell particles only.

We conclude with a few remarks on possible inclusion of
correlations in BUU calculations.  An important concept here is
the
time associated with the off-shell propagation.  In~scattering,
a time delay $\tau$ may be associated with
the propagation of the nucleon given by
$\tau=\hbar\Gamma/((\Delta E)^2 +
\Gamma^2/4)$ \cite{dan95,Me}.  The~$\tau$ in the case we
considered
is very short, which would allow the effects to be treated as an
effective three-body interaction.  A phenomenological model to
introduce such many particle interactions into calculations has
already been described in
Ref.\cite{bat92}.

This work was initiated at the Institute for Nuclear Physics.
We~thank U.~Mosel for conversations.  Support for the
work was given by the National Science Foundation under
Grant No.~PHY-9403666 and by the Department of Energy under
Grant No.~DE-FG06-90ER40561.

\begin{figure}
\caption{The graph for heavy particle production below
the two-body threshold. }
\end{figure}
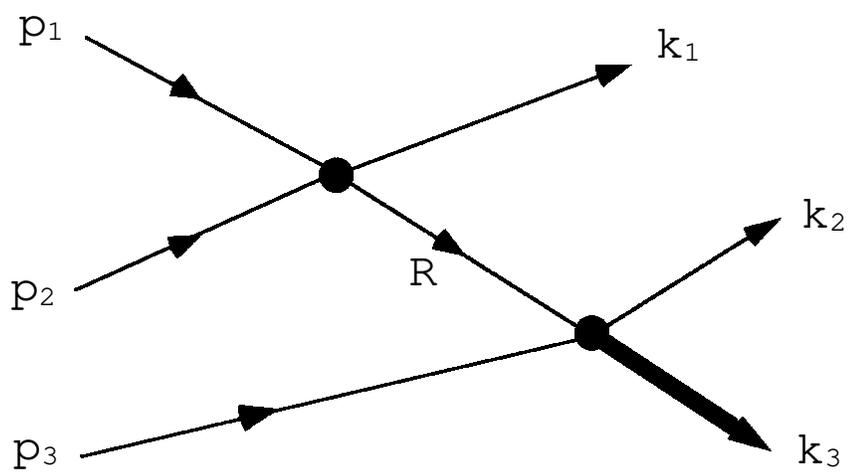
\end{document}